\pdfoutput=1   % needed for arXiv, and should be within the first 5 lines of the file!!!
\documentclass[usenatbib,usegraphicx,fleqn]{ctextemp_mn2e} % two columns, single spaced
\usepackage{amsmath,amssymb,bm,color,multirow,multicol,eufrak,relsize}
\usepackage{charter}   % better font
\newcommand{\fig}[1]{Fig. \ref{#1}}

\newcommand{\slfrac}[2]{\left.#1\middle/#2\right.}

\usepackage{ulem}
\title[Effect of an azimuthal magnetic field on a migrating planet]{The effect of an ordered azimuthal magnetic field on a migrating planet in a non-turbulent disc}

\author[M. L. Comins et al.]
{M.L. Comins$^1$\thanks{e-mail:mcomins@astro.cornell.edu},
M.M. Romanova$^1$, A.V. Koldoba$^{3,4}$, G.V. Ustyugova$^{3}$, R.V.E. Lovelace$^{1,2}$\\
$^1$ Department of Astronomy, Cornell University, Ithaca, NY 14853-6801, USA\\
$^2$ School of Applied \& Engineering Physics, Cornell University, Ithaca, NY 14853-6801, USA\\
$^3$ Keldysh Institute for Applied Mathematics, Moscow, Russia \\
$^4$ Moscow Institute of Physics and Technology, Dolgoprudny, Moscow Region, 141700, Russia}

\begin{document}

\maketitle

\begin{abstract}
In this work, we consider the physics of the interaction between a planet and
a magnetized gaseous protoplanetary disc.
We investigate the migration of a planet in a disc that is threaded with
an azimuthal magnetic field. We find that, for a larger magnetic field amplitude, there 
is an increasingly large positive torque on the planet from the disc, resulting in slowed 
and even outward migration. Our results indicate that magnetic resonances 
due to a purely azimuthal, ordered magnetic field can slow or stop the inward migration of 
Jupiter-mass, Saturn-mass, and $5 M_{\oplus}$ planets. 

\end{abstract}


\section{Introduction}\label{section:intro}
There are three primary types of planet migration within a gaseous disc \citep{ward1997}. 
Type I migration occurs when a planet remains embedded in
and exchanges angular momentum with the disc, and waves propagate 
through the disc as a result. 
Type II migration occurs when a planet is 
massive enough to open a gap in the disc, because the planet's 
Hill radius exceeds the scale height of the disc. The Hill radius is defined as
\begin{equation}
R_{\rm H} = r_{\rm p}\left(\frac{M_{\rm p}}{3M_{\star}}\right)^{1/3},
\end{equation}
where $M_{\star}$ is the mass of the central star, $M_{\rm p}$ is the mass
of the planet, and $r_{\rm p}$ is the orbital radius of the planet.
In type III migration, material passes through the corotation 
region, exchanging angular momentum with the planet and causing runaway migration of 
the planet in the process. If the planet initially moves inward, the runaway migration
direction is inward (and vice versa). 

\subsection{Hydrodynamic torques}
The planet revolves about the star in a circular orbit with period $P = \slfrac{2\pi}{\Omega_{\rm p}}$, 
where $\Omega_{\rm p}$ is the angular orbital frequency of the planet.
Spiral density waves are excited at the disc's Lindblad resonances when the planet 
tidally interacts with the disc. The angular momentum is redistributed in the disc after 
the planet ``dumps'' angular momentum at the resonances; the disc then exerts a 
gravitational torque back onto the planet, consequently changing the planet's orbital elements.

The planet excites $m$-armed waves in the disc with ``orbital'' frequencies
\begin{equation}
	\omega = m\Omega_{\rm p}.
\end{equation}
The dispersion relation for the waves in a low-temperature, low-mass disc is
\begin{equation}
	\left(\omega - m\Omega\right)^{2} = \kappa^{2},
\end{equation}
where $\Omega = \Omega(r)$ is the angular frequency of the disc rotation at radius $r$, 
$\omega - m\Omega$ is the Doppler shifted frequency of the $m$-armed wave as seen by an observer orbiting at the
disc's angular velocity, and $\kappa$ is the epicyclic frequency, which in a Keplerian disc is $\kappa = \Omega_{\rm Kep}$.
In other words,
\begin{equation}
	m^{2}(\Omega-\Omega_{\rm p})^{2} = \kappa^{2}.
\end{equation}
Substituting $\Omega(r) =  \sqrt{\slfrac{GM_{\star}}{{r}^{3}}}$ and $\Omega_{\rm p} = \sqrt{\slfrac{GM_{\star}}{r_{\rm p}^{3}}}$,
the locations of the Lindblad resonances are
\begin{equation}
	r_{\rm LR} = r_{\rm p} \left( \frac{m \pm 1}{m} \right)^{2/3}.
\end{equation}

For each value of $m$ (except $m=1$), there is one resonance located closer to the star
than the planet ($r_{\rm ILR} < r_{\rm p}$), known as the ``inner Lindblad
resonance'' (ILR), and one located farther from the star than the planet
($r_{\rm OLR} > r_{\rm p}$), known as the ``outer Lindblad resonance'' (OLR).
The parts of the disc interior to the planet exert a positive torque on the
planet and ``push'' the planet outward, while the parts of the disc exterior 
to the planet exert a negative torque on the planet and push the planet inward.

The torques exerted on the planet from each ILR and OLR are different in
magnitude and opposite in sign; the sum of these torques is referred to as
the ``differential Lindblad torque.'' The sign of the differential
Lindblad torque indicates the direction in which the planet will migrate. 
The differential Lindblad torque on the disc from the planet, as calculated by \citet{ttw2002}, is
\begin{equation}
	\Gamma_{\rm total} \propto \left(\frac{M_{\rm p} r_{\rm p} \Omega_{\rm p}}{c_{s} M_{\star}}\right)^{2} \Sigma(r_{\rm p}) r_{\rm p}^{4} \Omega_{\rm p}^{2} > 0,
\end{equation}
where $\Sigma(r_{\rm p})$ and $c_{s}$ are the surface density and isothermal sound speed in the disc at the orbital radius of the planet, respectively.
This implies that the torque on the planet from the disc is negative, $\Gamma_{\rm total} < 0$, 
resulting in inward migration of the planet as it loses angular momentum to the disc.

There is also a torque from the material corotating with the planet, of which
two primary contributors are the entropy- and vortensity-related torques,
where vortensity is defined as the ratio of the local vorticity to the surface density in the disc:
\begin{equation}
	\eta = \frac{\nabla \times \mathbf{v}}{\Sigma}.
\end{equation}
The corotation torque connected with vortensity is proportional to gradient of the vortensity across the corotation region
about the planet \citep{masset2001}. The corotation torque connected with the entropy is, analogously, proportional to the gradient of the
entropy across this same region \citep{paardekooper_etal_2010}. Both of these corotation torques can 
exert a positive torque on the planet, and the (positive) magnitude of the total corotation torques can exceed the 
magnitude of the (negative) differential Lindblad torque, possibly resulting in overall outward planet migration.
The specific details of these corotation torques are outside the scope of this paper, but see, e.g., 
\citet{pm2006,bm2008,pp2008,kbk2009,mc2009,mc2010,paardekooper_etal_2010,pbk2011} for more details.

\subsection{MHD torque}
When there exists a magnetic field in the disc, magnetic resonances can appear
that can also exert a significant torque on the planet \citep[e.g.,][]{terquem2003,ftn2005,fl2011}.
\citet{terquem2003} calculated the linear torque
on a planet in a fixed circular orbit and found that magnetic resonances can be 
generated near the planet, and that tightly wound 
waves can be generated near these resonances. The frequency of the waves excited by
the planet is equal to the Doppler-shifted frequency of a slow MHD wave that is 
propagating along the field lines in the disc \citep{terquem2003},
\begin{equation}
	m^{2}(\Omega-\Omega_{\rm p})^{2} = \frac{m^{2}c_{s}^{2}v_{\rm A}^{2}}{r^{2}(v_{A}^{2}+c_{s}^{2})}.
\end{equation}
Here, $m$ cancels out and $v_{\rm A}$ is the Alfv\'{e}n speed, given by \citet{terquem2003} as
\begin{equation}
	v_{\rm A}^{2} = \frac{\langle B^{2}\rangle}{\sqrt{\mu_{0}\Sigma}},
\end{equation} 
where $\langle B^{2} \langle$ is the vertically-integrated square of the magnetic field.
The locations of the inner magnetic resonance (IMR) and outer magnetic resonance (OMR), according to 
\citet{terquem2003}, are, respectively,
\begin{equation}
	\mathlarger{r_{\rm IMR} = r_{\rm p} - \frac{2 c_{s}}{3\Omega}\frac{1}{\sqrt{1 + \slfrac{v_{\rm A}^{2}}{c_{s}^{2}}}}}
	\label{terquem_rIMR}
\end{equation}
and
\begin{equation}
	\mathlarger{r_{\rm OMR} = r_{\rm p} + \frac{2 c_{s}}{3\Omega}\frac{1}{\sqrt{1 + \slfrac{v_{\rm A}^{2}}{c_{s}^{2}}}}},
	\label{terquem_rOMR}
\end{equation}
respectively. Here, $v_{A}$, $c_{s}$, and $\Omega$ are again evaluated at the location of the planet.
The waves associated with these resonances can propagate
interior and exterior to the Lindblad resonances, as well as between
the resonances.

The ratio between the matter and magnetic pressure in the disc is $\beta \equiv \slfrac{c_{s}^{2}}{v_{A}^{2}}$.
If the magnitude of $\beta$ near the planet (i.e., between the
magnetic resonances) is low enough in magnitude, the waves associated with the magnetic resonances 
can exert a positive torque that is larger in magnitude than the 
differential Lindblad torque.

Magnetic resonances can be important to planet migration in any part of the disc in which a
magnetic field is present. This includes the region of the disc in which planets are thought to form
(i.e., $1 - 5$ AU), as well as distances less than $0.1$ AU, which is near the stellar magnetosphere, 
where the field lines of the external magnetosphere can produce a significant azimuthal component 
of the field due to the differential rotation in the disc.

\citet{ftn2005} performed simulations of both a planet in a fixed
circular orbit and one that is allowed to migrate and
found, using 2D MHD simulations in a non-turbulent disc 
that the inward migration of a low-mass planet ($5 M_{\oplus}$) can be
reversed by the torque from the magnetic resonances.
A more recent paper by \citet{gbp2013} studied the effects of an MHD corotation torque
on a similarly low-mass planet in a 2D laminar disc with a weak azimuthal field threading the disc. 
The field was not strong enough to generate an appreciable torque from magnetic resonances, and
it was not strong enough to dominate the hydrodynamic corotation torques, but a ``torque excess'' 
attributed to the presence of the magnetic field was found. Their results correspond to the work done by
\citet{baruteau_etal_2011} and \citet{uribe_etal_2011}, which showed the existence of additional MHD
corotation torques in MRI-turbulent discs using 3D MHD simulations. 

The aim of this paper is to study how the torque exerted on the planet by magnetic resonances affects 
the migration of a planet in a non-turbulent two-dimensional disc threaded by an initially 
azimuthal magnetic field. We study how the 
total gravitational torque and semimajor axis evolves in cases of different
planet masses and the magnetic field amplitudes, while the disc mass is kept constant.

The plan of this paper is as follows. In \S \ref{section:model} we describe our physical model
and numerical setup. In \S \ref{section:sigma_beta_changes} we discuss the impact of a magnetic 
field on the surface density of the disc nearby the planet for several different planet masses. 
In \S \ref{section:torque_changes}, we discuss the effects of a magnetic
field on the total torque on the planet over time. 
In \S \ref{section:semimajor_axis_change}
we describe the impact of a magnetic field on the semimajor axis change for several planet masses.
In \S \ref{section:dimensional_units}, we discuss converting our data into dimensional units,
including the migration times of our simulated planets.
Finally, in \S \ref{section:conclusions}, we discuss
our conclusions.

%%%%%%%%%%%%%%%%%%%%%%%%%%%%%%%%%%%%%%%%%%%%%%%%%%%%%%%%%%%%%%%%%%%
%
%  MODEL
%
%%%%%%%%%%%%%%%%%%%%%%%%%%%%%%%%%%%%%%%%%%%%%%%%%%%%%%%%%%%%%%%%%%%

\section{Model}\label{section:model}
\subsection{MHD Equations}\label{theory:mhd-equations}
We utilize the MHD equations to numerically evaluate the perturbative effect of the planet on the disc:
\begin{enumerate}
  \item Continuity equation (conservation of mass)
	\begin{equation}
		\frac{\partial\Sigma}{\partial t} + \frac{1}{r}\frac{\partial}{\partial r}(r\Sigma v_{r}) + \frac{1}{r}\frac{\partial}{\partial\varphi}(\Sigma v_{\varphi}) = 0,
	\end{equation}
  where $\Sigma=\int\rho dz$ is the surface density (with $\rho$ the volume density), and $v_{r}$ and $v_{\varphi}$ are the 
  radial and azimuthal velocities, respectively.

  \item Radial equation of motion (conservation of momentum)
	\begin{eqnarray}
		\nonumber \frac{\partial}{\partial t}(\Sigma v_{r}) &+& \frac{1}{r}\frac{\partial}{\partial r}\left[r\left(\Sigma v_{r}^{2} + \Pi + \frac{\Psi_{rr}+\Psi_{\varphi\varphi}}{8\pi} - \frac{\Psi_{rr}}{4\pi}\right)\right] \\
		\nonumber 						       &+& \frac{1}{r}\frac{\partial}{\partial\varphi}\left(\Sigma v_{r}v_{\varphi}-\frac{\Psi_{r\varphi}}{4\pi}\right) \\
									       &=& \frac{\Pi}{r} +\frac{\Psi_{rr} + \Psi_{\varphi\varphi}}{8\pi r} - \Sigma\frac{GM_{\star}}{r^{2}} + \Sigma w_{r},  
	\label{eqn:radial_motion}
  	\end{eqnarray}
  where $\Pi = \int p dz$ is the surface pressure (with $p$ the volume pressure), and $\Sigma w_{r}$ is the radial force exerted on 
  the disc by the planet (per unit area of the disc), and $\Psi_{rr}$, $\Psi_{r\varphi}$, and $\Psi_{\varphi\varphi}$ are magnetic surface variables
  which are discussed in more detail in \S \ref{subsection:mag_variables}.

  \item Azimuthal equation of motion (conservation of angular momentum)
	\begin{eqnarray}
		\nonumber \frac{\partial}{\partial t}(\Sigma v_{\varphi}) &+& \frac{1}{r^{2}}\frac{\partial}{\partial r}\left[r^{2}\left(\Sigma v_{r}v_{\varphi}-\frac{\Psi_{r\varphi}}{4\pi}\right)\right] \\
		\nonumber 							      &+& \frac{1}{r}\frac{\partial}{\partial\varphi}\left(\Sigma v_{\varphi}^{2} + \Pi + \frac{\Psi_{rr}+\Psi_{\varphi\varphi}}{8\pi} - \frac{\Psi_{\varphi\varphi}}{4\pi}\right) \\
										      &=& \Sigma w_{\varphi},
	\label{eqn:az_motion}
	\end{eqnarray}
  where $\Sigma w_{\varphi}$ is the azimuthal force exerted on the disc by the planet (per unit
  area of the disc).

  \item Radial induction equation
  \begin{equation}
     \frac{\partial\Phi_{r}}{\partial t} + \frac{1}{r}\frac{\partial}{\partial\varphi}\left(v_{\varphi}\Phi_{r} - v_{r}\Phi_{\varphi}\right) = 0,
	\label{eqn:radial_induction}
  \end{equation}
  where $\Phi_{r}$ and $\Phi_{\varphi}$ are different magnetic surface variables, also discussed in \S \ref{subsection:mag_variables}.

  \item Azimuthal induction equation
  \begin{equation}
    \frac{\partial\Phi_{\varphi}}{\partial t}+ \frac{\partial}{\partial r}\left(v_{r}\Phi_{\varphi} - v_{\varphi}\Phi_{r}\right) = 0.
	\label{eqn:az_induction}
  \end{equation}

  \item Entropy balance equation
	\begin{equation}
		\frac{\partial}{\partial t}\left[\Sigma K\right] + \frac{1}{r}\frac{\partial}{\partial r}\left[r\Sigma K v_{r}\right] 
                                                                                                  + \frac{1}{r}\frac{\partial}{\partial\varphi}(\Sigma K v_{\varphi}) = 0,
	\end{equation}
where $K = \slfrac{\Pi}{\Sigma^{\gamma}}$ is a function analogous to entropy, 
and $\gamma = 5/3$ is the adiabatic index for a monatomic ideal gas.
\end{enumerate}

Moreover, following the $\alpha$ prescription of \citet{ss1973}, we use a very small 
viscosity ($\alpha = 0.001$) to isolate the effects of slow magnetosonic waves in the disc
by smoothing out the effects of fast magnetosonic waves in the disc.

\subsubsection{Magnetic surface variables}\label{subsection:mag_variables}
The ``volume'' values for the radial and azimuthal magnetic fields are given by $B_{r}$ and $B_{\varphi}$. 

The usage of surface values for the variables related to the magnetic field are not as easily defined and
used as $\Sigma$ and $\Pi$, and so we describe them in detail here. 
The ``volume'' values for the radial and azimuthal magnetic fields are given by $B_{r}$ and $B_{\varphi}$. 
The vertically-integrated fields may be defined as $\Phi_{r} = \int B_{r} dz$ and $\Phi_{\varphi} = \int B_{\varphi} dz$,
 as done with $\Sigma$ and $\Pi$. There are also terms involving magnetic flux or energy that involve products of these variables: 
$B_{r}^{2}$, $B_{\varphi}^{2}$, and $B_{r}B_{\varphi}$. We define the following vertically-integrated quantities: $\Psi_{rr}~=~\int B_{r}^{2} dz$,
$\Psi_{r\varphi} = \int B_{r}B_{\varphi} dz$, and $\Psi_{\varphi\varphi} = \int B_{\varphi}^{2} dz$.

As shown above, the induction equations use $\Phi_{r}$ and $\Phi_{\varphi}$,
while the equations of motion use $\Psi_{rr}$, $\Psi_{r\varphi}$, and $\Psi_{\varphi\varphi}$.
As such, we need a way to relate $\Phi$ and $\Psi$. We can do this using a ``magnetic'' thickness of the disc, 
$H$. Using the definitions of $\Phi$ and $\Psi$, with $H$, we find that
\begin{align}
\Psi_{rr} &= \frac{\Phi_{r}\Phi_{r}}{H}; & \Psi_{r\varphi} &= \frac{\Phi_{r}\Phi_{\varphi}}{H}; & \Psi_{\varphi\varphi} &= \frac{\Phi_{\varphi}\Phi_{\varphi}}{H}.
\label{eqn:Psi_Phi}
\end{align}
We suggest that the coefficient $H$ is the same in all three relations.

By relating $\Phi$ and $\Psi$ in this way, we can define one more variable related to the magnetic
field such that the MHD equations are parameterized with respect to a single magnetic field variable,
\begin{align}
	\mathfrak{B}_{r} &= \frac{\Phi_{r}}{\sqrt{H}}; & \mathfrak{B}_{\varphi} &= \frac{\Phi_{\varphi}}{\sqrt{H}}.
	\label{eqn:F_Phi}
\end{align}
Then, the magnetic terms in Equations (\ref{eqn:radial_motion}) - (\ref{eqn:az_induction}) take their usual form. The radial equation of motion becomes
	\begin{eqnarray}
		\nonumber \frac{\partial}{\partial t}(\Sigma v_{r}) &
		+& \frac{1}{r}\frac{\partial}{\partial r}\left[r\left(\Sigma v_{r}^{2} 
		+ \Pi 
		+ \frac{\mathfrak{B}_{r}^{2}+\mathfrak{B}_{\varphi}^{2}}{8\pi} - \frac{\mathfrak{B}_{r}^{2}}{4\pi}\right)\right] \nonumber \\
		&+& \frac{1}{r}\frac{\partial}{\partial\varphi}\left(\Sigma v_{r}v_{\varphi}-\frac{\mathfrak{B}_{r}\mathfrak{B}_{\varphi}}{4\pi}\right) \nonumber \\
		&=& \frac{\Sigma v_{\varphi}^{2}}{r} 
		+ \frac{\Pi}{r} 
		+\frac{\mathfrak{B}_{r}^{2} + \mathfrak{B}_{\varphi}^{2}}{8\pi r} \nonumber \\
		&-& \Sigma\frac{GM_{\star}}{r^{2}} + \Sigma w_{r},
  	\end{eqnarray}
  and the azimuthal equation of motion becomes
	\begin{eqnarray}
		\frac{\partial}{\partial t}(\Sigma v_{\varphi}) 
		&+& \frac{1}{r^{2}}\frac{\partial}{\partial r}\left[r^{2}\left(\Sigma v_{r}v_{\varphi}-\frac{\mathfrak{B}_{r}\mathfrak{B}_{\varphi}}{4\pi}\right)\right] \nonumber \\
		&+& \frac{1}{r}\frac{\partial}{\partial\varphi}
		\left(\Sigma v_{\varphi}^{2} + \Pi + \frac{\mathfrak{B}_{r}^{2}+\mathfrak{B}_{\varphi}^{2}}{8\pi} - \frac{\mathfrak{B}_{\varphi}^{2}}{4\pi}\right) \nonumber \\
	          &=& \Sigma w_{\varphi}.
	\end{eqnarray}
  For simplicity, we accept that $H = {\rm const}$. Then, the radial and azimuthal induction equations become, respectively,
\begin{equation}
	\frac{\partial \mathfrak{B}_{r}}{\partial t} + \frac{1}{r}\frac{\partial}{\partial\varphi}\left(v_{\varphi}\mathfrak{B}_{r} - v_{r}\mathfrak{B}_{\varphi}\right) = 0.
\end{equation}
and
\begin{equation}
	\frac{\partial \mathfrak{B}_{\varphi}}{\partial t}+ \frac{\partial}{\partial r}\left(v_{r}\mathfrak{B}_{\varphi} - v_{\varphi}\mathfrak{B}_{r}\right) = 0.
\end{equation}

\subsection{Planetary equation of motion}
We calculate the equations of motion in the stellar reference frame. This system is not inertial, because
the star (due to the gravitational influence from the planet and disc) also revolves about the center of mass of the system.
Thus, the inertial force is added to the equation of motion for both the disc and the planet.
Assuming that the inertial acceleration is only due to the gravitational attraction between the star and the planet
(but not the disc), the inertial force per unit mass (i.e., acceleration) is
\begin{equation}
	\mathbf{w}_{i} = -\frac{GM_{\rm p}}{r_{\rm p}^{3}}\mathbf{r}_{\rm p}.
\end{equation}
The total acceleration is then
\begin{equation}
	\mathbf{w} = \mathbf{w}_{\rm p} + \mathbf{w}_{i},
\end{equation}
which is converted into polar coordinates to obtain $w_{r}$ and $w_{\varphi}$.

We use the same gravitational potential as that used in the three-dimensional simulations in \citet{kbk2009} and \citet{kk2006},
\begin{equation}
 \Phi = -\frac{GM_{\star}}{r} + \Phi_{\rm p},
\end{equation}
where $\Phi_{\rm p}$ reflects the gravitational influence of the planet on the disc,
\begin{equation}
\Phi_{\rm p} = 
  \begin{cases}
    -\frac{GM_{\rm p}}{d}  \left[\left(\frac{d}{r_{\rm sm}}\right)^{4}
        -2\left(\frac{d}{r_{\rm sm}}\right)^{3}
        +2\frac{d}{r_{\rm sm}}
      \right]
     & \text{for } d\leq r_{\rm sm} \\
    -\frac{GM_{\rm p}}{d} & \text{for } d > r_{\rm sm},
  \end{cases}
\end{equation}
with $d = \lvert {\mathbf r} - {\mathbf r}_{\rm p}\rvert$ being the distance between
the planet and a point in the disc, and $r_{\rm sm}$ the
smoothing radius. We use a smoothing radius of $r_{\rm sm}=0.8R_{\rm H}$, where
$R_{\rm H}$ is the Hill radius. Although our simulations are two-dimensional,
we use this potential in anticipation of future three-dimensional work.

We use this potential to calculate the force from the planet on each 
of the fluid elements in the disc (per unit mass):
\begin{equation}
	\mathbf{w}_{\rm p} = -\nabla\Phi_{\rm p} = 
	\begin{cases}
		-\frac{GM_{\rm p}}{r_{\rm sm}^{3}} \left( \frac{3d}{r_{\rm sm}} - 4\right)\mathbf{d} & \text{for $d \leq r_{\rm sm}$} \\
		-\frac{GM_{\rm p}}{d^{3}}\mathbf{d}							    & \text{for $d > r_{\rm sm}$}.
	\end{cases}
\end{equation}
The force exerted {\it on} the planet by a particular fluid element is the 
acceleration with opposite sign, multiplied by the mass of the fluid element,
\begin{equation}
	d\mathbf{f} = -dM\mathbf{w} = dM(\nabla\Phi_{\rm p}-\mathbf{w}_{i}),
	\label{eqn:df}
\end{equation}
where $dM = \Sigma rdrd\varphi$.
We then calculate the total force exerted on the planet by the disc by
integrating over the disc within the computational domain,
\begin{equation}
	\mathbf{F}_{\rm disc\rightarrow p} = \int_{\rm disc} \psi d\mathbf{f}
                                                                    = \int_{\rm disc} \psi (\mathbf{w}_{i} - \nabla\Phi_{\rm p})dM,
\end{equation}
where $\psi$ is a tapering function used to exclude the inner parts of the planet's Hill sphere from the force \citep{kbk2009}:
\begin{equation}
	\psi = \left[\exp\left({\frac{r_{\rm sm}-d}{0.1r_{\rm sm}}}\right)+1\right]^{-1}.
\end{equation}
We use the planet's equation of motion to find the position, $\mathbf{r}_{\rm p}$, and velocity, $\mathbf{v}_{\rm p}$, of the planet at each time step:
\begin{equation}
	M_{p} \frac{d\mathbf{v}_{\rm p}}{dt} = -\frac{GM_{\star}M_{\rm p}}{r_{\rm p}^{3}}\mathbf{r}_{\rm p} 
                                                                             -\frac{GM_{\rm p}^{2}}{r_{\rm p}^{3}}\mathbf{r}_{\rm p} + \mathbf{F}_{\rm disc\rightarrow p}.
\end{equation}

In addition to this, we calculate the planet's orbital semimajor axis, $a$, and eccentricity, $e$, to
describe the general properties of its orbit. These orbital elements were calculated via the planet's orbital energy
and angular momentum per unit mass \citep{md1999}:
\begin{align}
	E              &= \frac{1}{2}\lvert{\mathbf v}_{\rm p}\rvert^{2} - \frac{GM_{\star}}{a},\\
	h &= \mathbf{r}_{\rm p} \times \dot{\mathbf{r}}_{\rm p},\\
	a &= -\frac{1}{2}\frac{GM_{\star}}{E},\\
\text{and}&\nonumber\\
	e &= \sqrt{1 - \frac{h^{2}}{GMa}}.
\end{align}

The gravitational torque in the $z$ direction on the planet is the sum over the
torques from each fluid element: $T_{z}~=~\int_{\rm disc} \left[\mathbf{r} \times \psi d\mathbf{f} \right]_{z}$. 
Effectively, torques due to the magnetic resonances alter the disc's surface 
density profile, which then alters the gravitational torque on the planet. 

\subsection{Reference units}\label{model:units}
Our code uses dimensionless variables; for each dimensionless quantity, $\widetilde{Q}$, the physical quantity, $Q$, is recovered via
\begin{equation}
	Q = Q_{0} \widetilde{Q},
	\label{eqn:dimensionless}
\end{equation}
where $Q_{0}$ is a ``reference quantity'' that converts the simulation variables into their dimensional counterparts. 
The value of $Q_{0}$ is chosen to reflect realistic astrophysical systems.

We choose the mass of the star, $M_{\star}$, and the reference distance, $r_{0}$. The reference
velocity is the Keplerian orbital velocity, $v_{0} = \sqrt{\slfrac{GM_{\star}}{r_{0}}}$. The reference time scale is
defined as $t_{0} = \slfrac{r_{0}}{v_{0}}$, and the reference rotational period is $P_{0} = 2\pi t_{0}$.
The inner and outer radii of the simulation region are given by $r_{\rm in} = r_{0}$ and $r_{\rm out} \approx 33 r_{0}$,
and a disc height typical for a protoplanetary disc is chosen ($H = 0.05 r_{0}$). 

We choose the mass of the planet relative to the mass of the star, $M_{\rm p} = \widetilde{m}_{\rm p}M_{\star}$.
We also introduce a characteristic mass of the disc, $M_{\rm d} = \widetilde{m}_{\rm d}M_{\star}$, in which $\widetilde{m}_{\rm d} = \kappa \widetilde{m}_{\rm p}$.
Because the disc mass can be altered via both $M_{\rm p}$ and $\kappa$, changing the planet mass requires a compensating
change in $\kappa$ in order to keep the mass of the disc constant. For $M_{\rm p} = M_{\rm Jup}$, $\widetilde{m}_{\rm p} = 10^{-3}$ and $\kappa = 1$.
Thus, because (approximately) $M_{\rm Sat} = 0.3 M_{\rm Jup}$, $\widetilde{m}_{\rm p} = 3\times 10^{-4}$ and $\kappa = 3$ for the case of a Saturn-mass planet.
Also, $5 M_{\oplus} = 0.015 M_{\rm Jup}$, and so $\widetilde{m}_{\rm p} = 1.5\times 10^{-5}$ and $\kappa = 66.67$ for the case of a 5 $M_{\oplus}$ planet.
The reference surface density is $\Sigma_{0} = \slfrac{M_{\rm d}}{r_{0}^{2}}$.

To find the characteristic reference value for the magnetic field, $B_{0}$, we utilize the previously-defined
surface magnetic field $\mathfrak{B}$,
\begin{equation}
	\mathfrak{B}_{0}^{2} = B_{0}^{2}r_{0}.
	\label{b_0_1}
\end{equation}
Furthermore, we can vertically integrate $B_{0}^{2} = \rho_{0} v_{0}^{2}$ to attain
\begin{equation}
	\mathfrak{B}_{0}^{2} = \Sigma_{0} v_{0}^{2}.
	\label{b_0_2}
\end{equation}
Combining Equations (\ref{b_0_1}) and (\ref{b_0_2}) yields $B_{0}$:
\begin{equation}
	B_{0} = \sqrt{\slfrac{\Sigma_{0} v_{0}^{2}}{r_{0}}}.
\end{equation}

The values for our reference quantities are shown in Table \ref{table:units}. Note that, because the 
migration time scale for an embedded planet is inversely proportional to the disc's surface density, 
we are able to reduce the computational time necessary to see migration by increasing the surface 
density; this is why we use a large surface density for $\Sigma_{0}$. For more details, please see 
\S \ref{section:dimensional_units}. 

\begin{table}
\begin{tabular} {c r l l}
\hline
Variable            &   Meaning                                              	 & Value                	       \\
\hline
$r_{0}$            &  Reference distance scale                    	 & $0.5$ AU                            	       \\
                         &                                                              	 & $7.48\times 10^{12}$ cm 	       \\
\hline
$M_{\star}$     & Stellar mass                                          	 & $1 M_{\odot}$			       \\
                         &                                                              	 & $1.989\times 10^{33}$ g               \\
\hline
$v_{0}$            & Reference (Keplerian) velocity            	 & $42.1$ km s$^{-1}$		         \\
\hline
$t_{0}$             & Reference timescale                            	 & $20.6$ days\\
\hline
$P_{0}$             & Reference (Keplerian) rotation period     	 & $129$ days\\
\hline
$M_{\rm d}$            & Reference (disc) mass                                   	 & $1.989\times 10^{30}$ g\\
\hline
$\Sigma_{0}$    & Reference disc surface density                   & $3.55\times 10^{4}$ g cm$^{-2}$\\
\hline
$B_{0}$             & Reference magnetic field strength              & $0.29$ kG\\
\hline
\end{tabular}
\caption{Reference units. See a more detailed description in \S \ref{model:units}.}
\label{table:units}
\end{table}

\subsection{Initial conditions}\label{num-model:ic-bc}

In order to accurately model this system, we must establish quasi-equilibrium initial conditions, such that 
the disc is approximately in mechanical equilibrium. Quasi-equilibrium between the rotation, pressure, and 
gravity of the system is achieved following a method similar to that described in, e.g., 
\S 2.4 of \citet{Romanova2002}, \citet{ukrl2006} and \citet{Dyda2013}.

\begin{figure}
	\centering
	\includegraphics[width=\columnwidth]{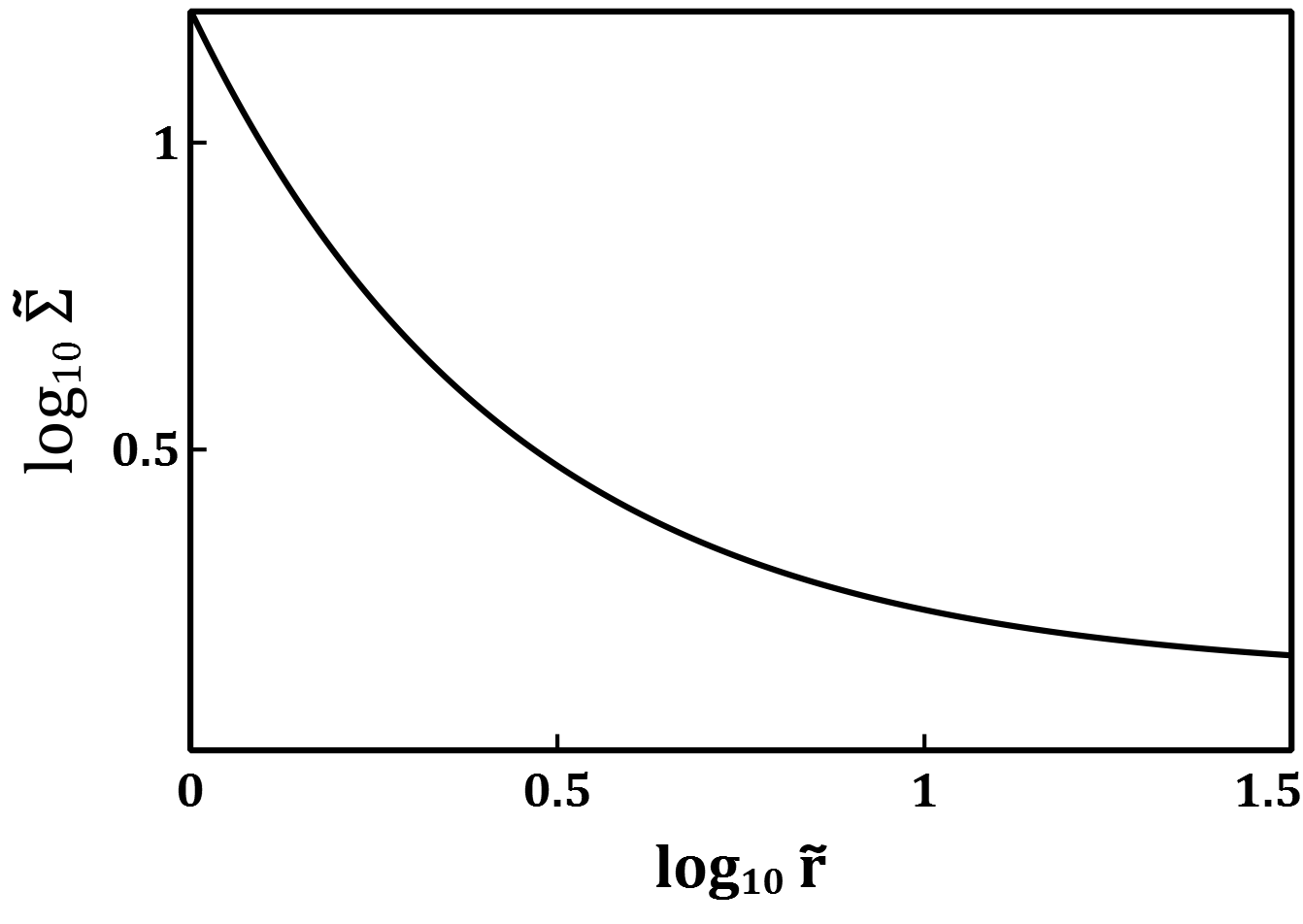}
	\caption{Initial radial surface density profile for all cases.}
	\label{fig:initial_1d_sigma}
\end{figure}

The initial pressure at $r = r_{0}$ is $\Pi_{0}$,
and the disc is initially barotropic, such that
\begin{equation}
	\Sigma(\Pi) = \frac{\Pi}{T_{\rm d}},
\end{equation} 
where $T_{\rm d}$ is the temperature in the disc\footnote{$\Pi_{0}$ does not denote a reference value here; it is instead defined as $\Pi_{0}\equiv\Pi(r_{\rm in}) = \Pi(r_{0})$.}.
This implies that the pressure can be calculated using Bernoulli's equation
\begin{equation}
	F + \Phi_{c} + \Phi = W = \text{constant}.
\end{equation}
Here $\Phi = -\slfrac{GM_{\star}}{r}$ is the gravitational potential,  
$\Phi_{c}$ is the centrifugal potential
\begin{equation}
	\Phi_{\rm c}(r) = k\frac{GM_{\star}}{r},
\end{equation}
where $k$ characterizes how different the disc is from Keplerian (i.e., our disc is slightly
sub-Keplerian in order to balance the pressure gradient, such that $k = 0.975$), and $F$ is
\begin{equation}
	F = \int_{\Pi_{0}}^{\Pi} {\frac{d\Pi}{\Sigma}} = T_{\rm d} \int_{\Pi_{0}}^{\Pi} {\frac{d\Pi}{\Pi}} = T_{\rm d} \ln \left(\frac{\Pi}{\Pi_{0}}\right).
	\label{eqn:F_init}
\end{equation}
At $r = r_{0}$, $F = 0$ and 
\begin{equation}
	W = \Phi(r_{0}) + \Phi_{\rm c}(r_{0}) = (k-1)\frac{GM_{\star}}{r_{0}}.
\end{equation}
By calculating $\Phi$ and $\Phi_{\rm c}$ throughout the disc, we can use $F = W - (\Phi + \Phi_{\rm c})$ 
to calculate the pressure distribution in the disc,
\begin{equation}
	\Pi = \Pi_{0} \exp\left(\frac{F}{T_{\rm d}}\right).
\end{equation}
Then, the initial density distribution is (shown in \fig{fig:initial_1d_sigma})
\begin{equation}
	\Sigma = \frac{\Pi}{T_{\rm d}} = \frac{\Pi_{0}}{T_{\rm d}} \exp\left(\frac{F}{T_{\rm d}}\right).
	\label{eqn:sigma_init}
\end{equation}

\begin{figure}
	\centering
	\includegraphics[width=\columnwidth]{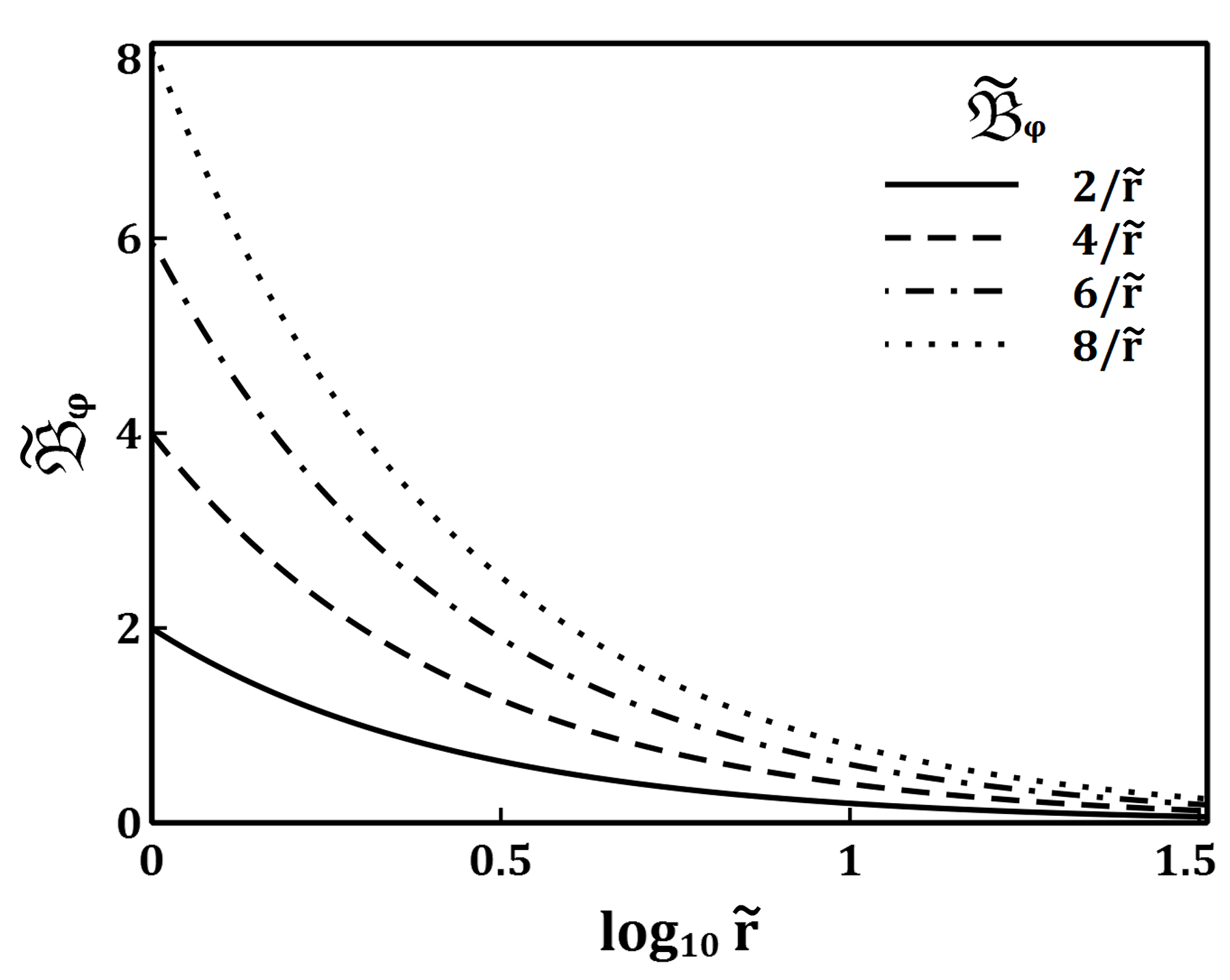}
	\caption{Initial radial profile of $\widetilde{\mathfrak{B}}_{\varphi}$ through the disc, for $\widetilde{\mathfrak{B}}_{\varphi} = \slfrac{2}{\widetilde{r}}$ (solid line),
		    $\widetilde{\mathfrak{B}}_{\varphi} = \slfrac{4}{\widetilde{r}}$ (dashed line), $\widetilde{\mathfrak{B}}_{\varphi} = \slfrac{6}{\widetilde{r}}$ (dash-dotted line), and
                         $\widetilde{\mathfrak{B}}_{\varphi} = \slfrac{8}{\widetilde{r}}$ (dotted line).}
	\label{fig:initial_1d_bf}
\end{figure}

\begin{figure}
  \centering
  \includegraphics[width=\columnwidth]{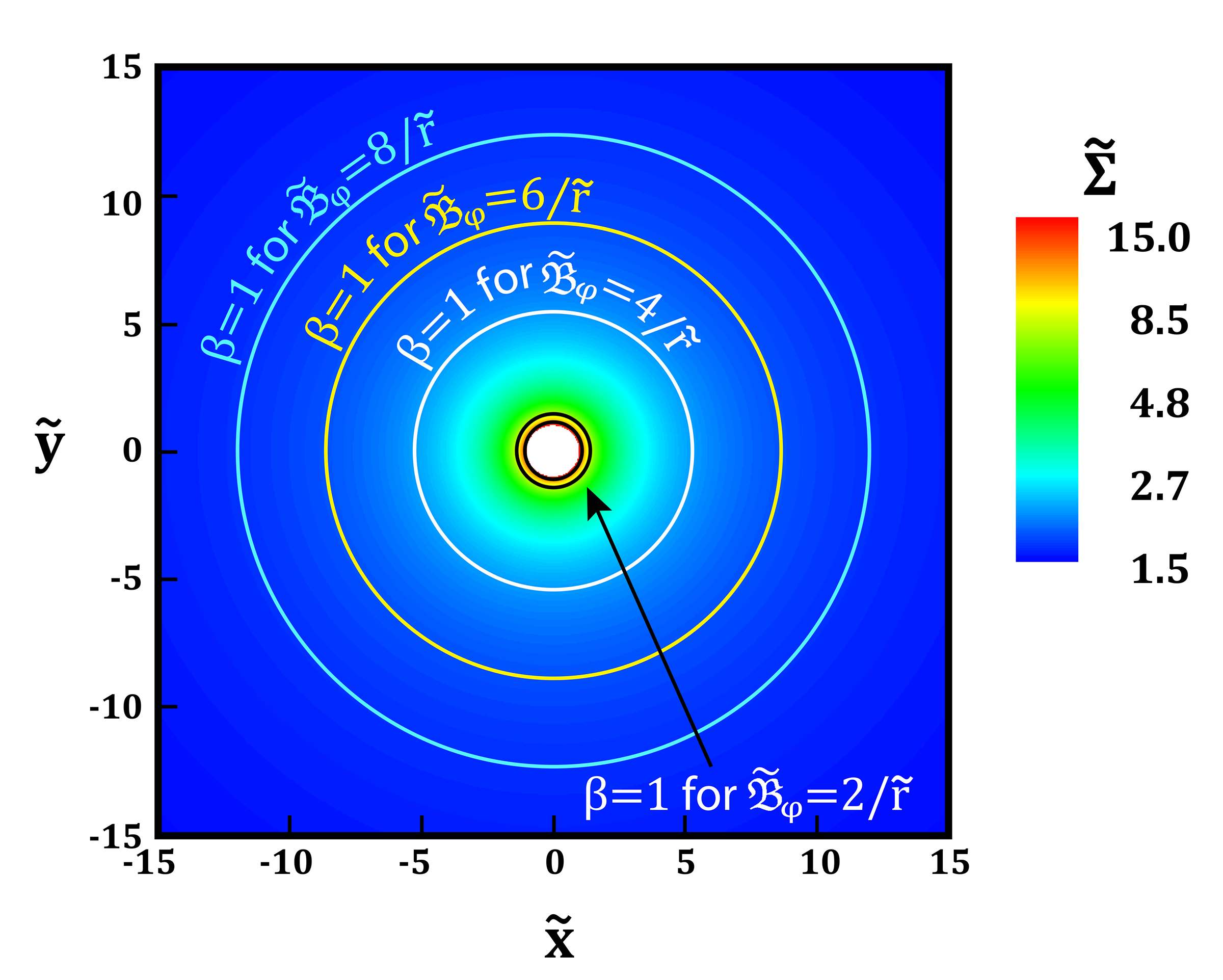}
  \caption{The initial surface density for all cases, with the location of the $\beta=1$ surface superimposed
                 on the disc for several magnetic field amplitudes: $\widetilde{\mathfrak{B}}_{\varphi} = \slfrac{2}{\widetilde{r}}$ (black),
		    $\widetilde{\mathfrak{B}}_{\varphi} = \slfrac{4}{\widetilde{r}}$ (white), $\widetilde{\mathfrak{B}}_{\varphi} = \slfrac{6}{\widetilde{r}}$ (yellow), and
                         $\widetilde{\mathfrak{B}}_{\varphi} = \slfrac{8}{\widetilde{r}}$ (cyan).}
  \label{fig:init_sigma_beta_xy}
\end{figure}

The magnetic field is initialized as purely toroidal via
\begin{equation}
	\mathfrak{B}_{\varphi} = \frac{b}{r} {\widehat{\boldsymbol{\mathrm{\varphi}}}}.
	\label{mag_field_init_eqn}
\end{equation}
Varying the magnetic field amplitude via $b$
effectively changes the initial location of the $\beta=1$ surface relative to
the planet.
In this work, we define
$\beta$ as
\begin{equation}
	\beta \equiv\frac{\Pi}{\slfrac{\mathfrak{B}^{2}}{8\pi}}.
\end{equation}
 The initial radial distribution of $\widetilde{\mathfrak{B}}_{\varphi}$ through the disc is shown in 
\fig{fig:initial_1d_bf} for several values of the field amplitude, and 
\fig{fig:init_sigma_beta_xy} shows the initial surface density for all cases, with
the $\beta=1$ lines shown for several different magnetic field amplitudes. 
The positions of the $\beta = 1$ lines relative to the initial orbital radius of the planet
(at $\widetilde{r} = 5$) indicate that we are in an intermediate to strong range of
magnetic field strengths.

\subsection{Grid and boundary conditions}\label{num-model:grid}

We use a uniform polar grid in the $r$ and $\varphi$ directions, with $N_{r}$ defined to be the number of radial grid cells
and $N_{\varphi}$ the number of azimuthal grid cells. 
We use a fixed boundary condition at the inner boundary and outflow boundary conditions 
at the outer boundary for the surface density and pressure, as well as for the surface magnetic
field and velocity components.

We apply a wave damping algorithm near the inner boundary similar to the method used in
Section 3.2 of \citet{dvb+2006} to avoid nonphysical wave reflections. We damp waves
after every time step for $\widetilde{r} < \widetilde{r}_{\rm damp}$, where $\widetilde{r}_{\rm damp} = 1.375$,
similar to the inner damping region chosen in \citet{dvb+2006}. We define a coefficient:
\begin{equation}
	\mathcal{J} = 1 + \frac{\Delta \widetilde{t}}{\widetilde{P}_{\rm orb}}\left( \frac{\widetilde{r}_{\rm damp}-\widetilde{r}}{\widetilde{r}_{\rm damp}-\widetilde{r}_{\rm in}}\right)^{2},
\end{equation}
where $\Delta\widetilde{t}$ is the current time step and $\widetilde{P}_{\rm orb}$ is the Keplerian orbital period at $\widetilde{r}$. 
Then, in the damping region (i.e., $\widetilde{r} < \widetilde{r}_{\rm damp}$), we apply the following damping conditions,
\begin{align}
	\widetilde{v}_{r,{\rm damp}}       &= \widetilde{v}_{r,{\rm init}} + \mathcal{J} (\widetilde{v}_{r} - \widetilde{v}_{r,{\rm init}}), \\
	\widetilde{v}_{\varphi,{\rm damp}} &= \widetilde{v}_{\varphi,{\rm init}} + \mathcal{J} (\widetilde{v}_{\varphi} - \widetilde{v}_{\varphi,{\rm init}}), 
\end{align}
and
\begin{equation}
	\widetilde{\Sigma}_{\rm damp} = \widetilde{\Sigma}_{\rm init} + \mathcal{J} (\widetilde{\Sigma} - \widetilde{\Sigma}_{\rm init}),
\end{equation}
where $\widetilde{v}_{r,{\rm init}} = 0$ and ${\widetilde{v}_{\varphi,{\rm init}} = \widetilde{v}_{\rm Kep}}$ are the (dimensionless)
initial radial and azimuthal velocities, respectively. Finally, the entropy is updated using the damped surface density value.

%%%%%%%%%%%%%%%%%%%%%%%%%%%%%%%%%%%%%%%%%%%%%%%%%%%%%%%%%%%%%%%%%%%
%
%  RESULTS
%
%%%%%%%%%%%%%%%%%%%%%%%%%%%%%%%%%%%%%%%%%%%%%%%%%%%%%%%%%%%%%%%%%%%

\section{Change in $\tilde{\Sigma}$ and $\beta$ near the planet}\label{section:sigma_beta_changes}

\begin{figure}
  \centering
  \includegraphics[width=\columnwidth]{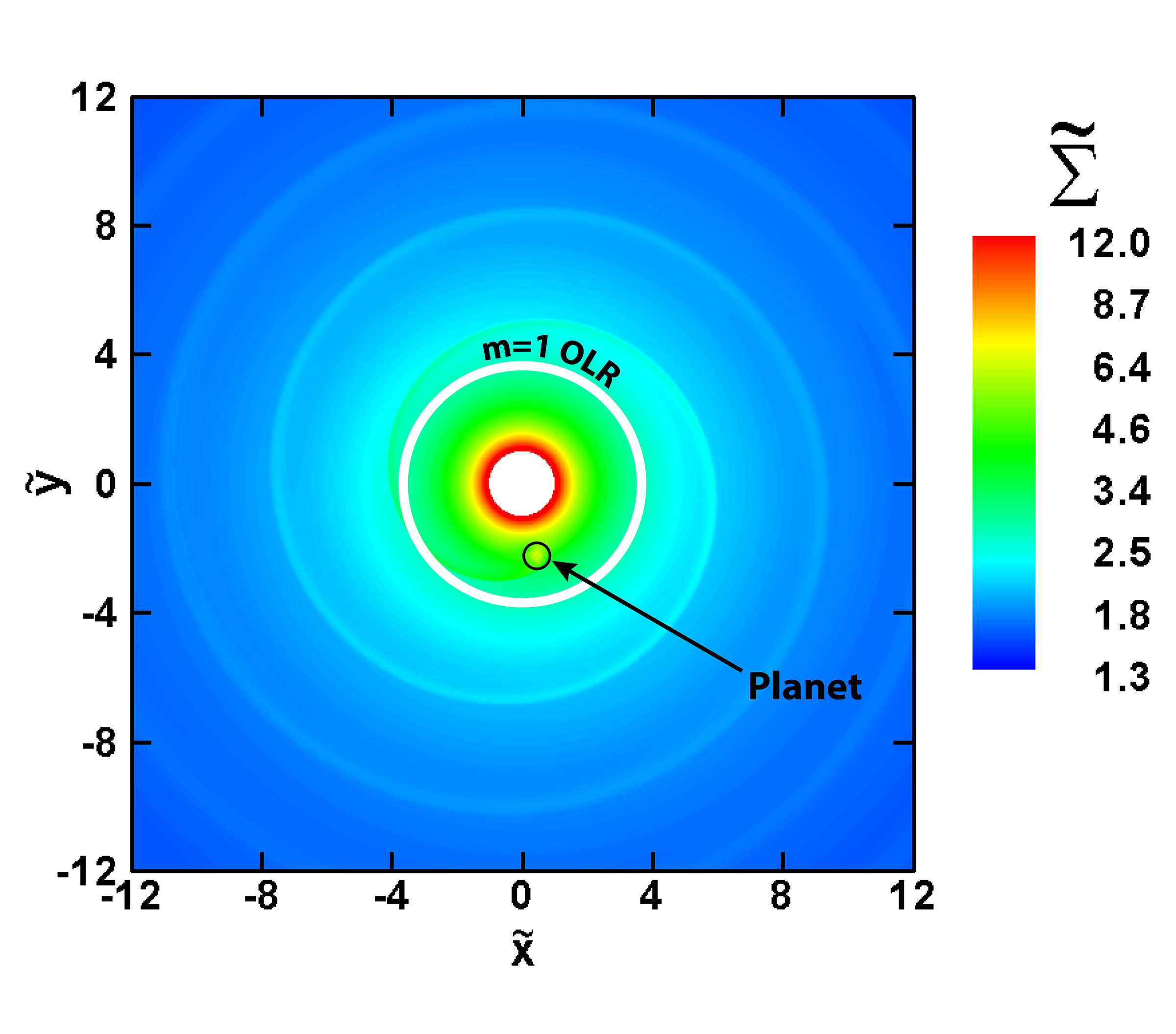}
  \caption{The surface density distribution for $\widetilde{\mathfrak{B}}_{\varphi} = 0$ (hydrodynamic case) at $\widetilde{t} = 100$
   for the case of a Jupiter-mass planet. The $m~=~1$ OLR location is labeled and shown by the 
  solid white circle; the planet's location is shown by the solid black circle of arbitrary size. Of the possible Lindblad resonances, 
  we choose to show only the $m=1$ OLR and the one-armed spiral wave excited by this resonance.}
  \label{fig:jup-lindblad-m1}
\end{figure}

\fig{fig:jup-lindblad-m1} shows the surface density distribution at $\widetilde{t} = 100$ for the hydrodynamic model in the case of a Jupiter-mass planet.
We only display out to $\widetilde{r} = 12$ in this figure to highlight variations in $\widetilde{\Sigma}$ more clearly.
The planet is indicated with a solid black circle of arbitrary size, and the location of the $m=1$ outer Lindblad resonance is labeled. 
The primary driver of migration in this case is the differential Lindblad torque, of which the torque from the $m=1$ OLR is the largest contributor.

\begin{figure*}
  \centering
  \includegraphics[width=\textwidth]{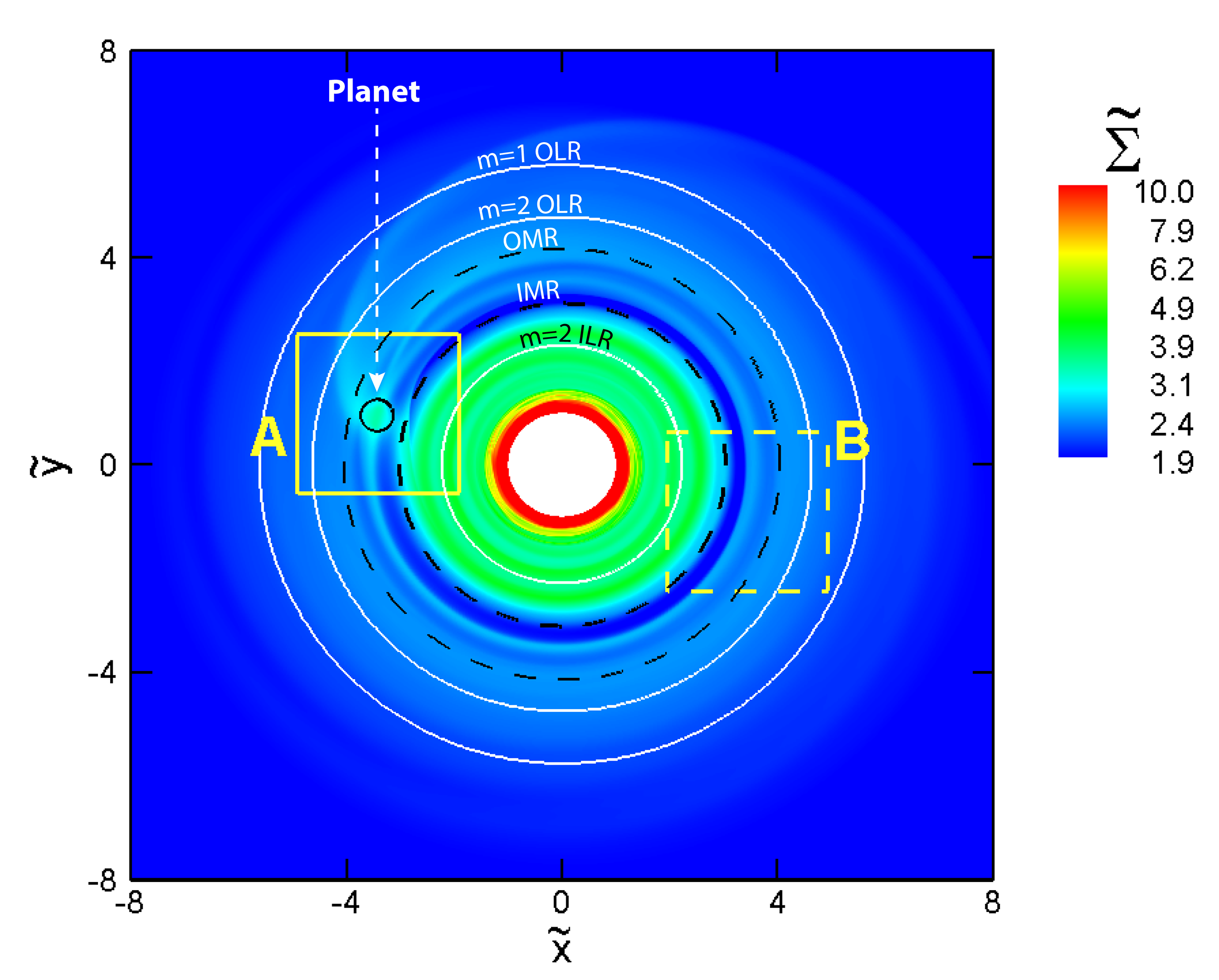}
  \caption{The surface density distribution at $\widetilde{t} = 600$ for $\widetilde{\mathfrak{B}}_{\varphi} = \slfrac{6}{\widetilde{r}}$
    for the case of a Jupiter-mass planet. The Lindblad resonances are marked by solid white circles, the magnetic resonances are
    marked by dashed black circles, and the planet's location is indicated by a solid black circle of arbitrary size.
    The two highlighted regions, A and B, are examined further in Figs. \ref{fig:jup-mhd-a} and \ref{fig:jup-mhd-b}.}
  \label{fig:jup-mhd-total}
\end{figure*}

\begin{figure*}
  \centering
  \includegraphics[width=\textwidth]{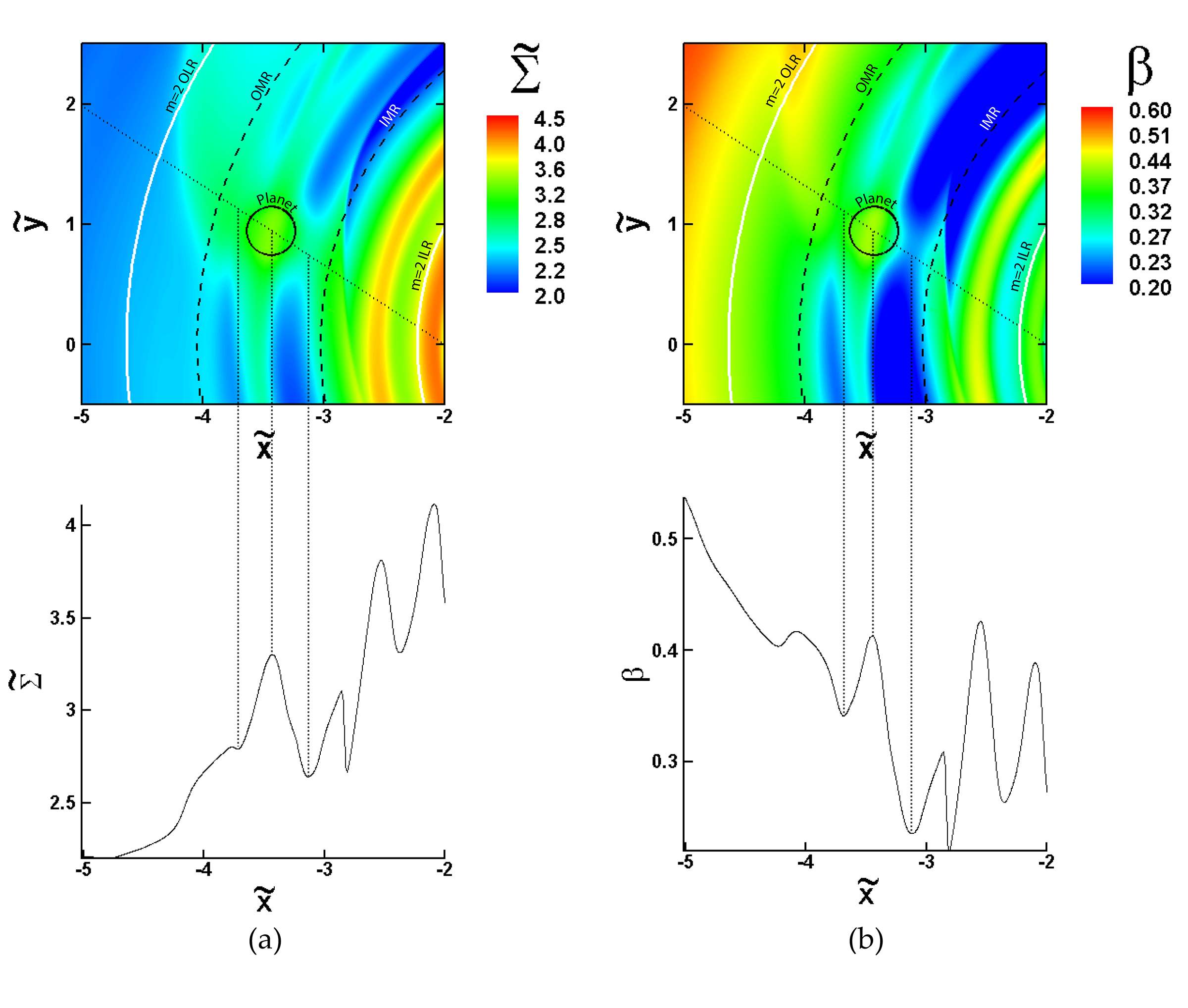}
  \caption{The (a) $\widetilde{\Sigma}$ and (b) $\beta$ distribution at $\widetilde{t} = 600$ for $\widetilde{\mathfrak{B}}_{\varphi} = \slfrac{6}{\widetilde{r}}$ 
   for a Jupiter-mass planet in region A indicated in \fig{fig:jup-mhd-total}. The IMR, OMR, $m=2$ OLR, and $m=2$ ILR are 
   labeled, as well as the location of the planet, in the top panel. $\widetilde{\Sigma}$ is sampled along a scanline indicated by a straight dotted line in the top panel and 
   plotted as a one-dimensional profile in the bottom panel of both (a) and (b). Vertical lines connecting the top and bottom panels 
   highlight variations in $\widetilde{\Sigma}$ and $\beta$ between the magnetic resonances and near the planet's orbital radius.}
  \label{fig:jup-mhd-a}
\end{figure*}

\begin{figure*}
  \centering
  \includegraphics[width=\textwidth]{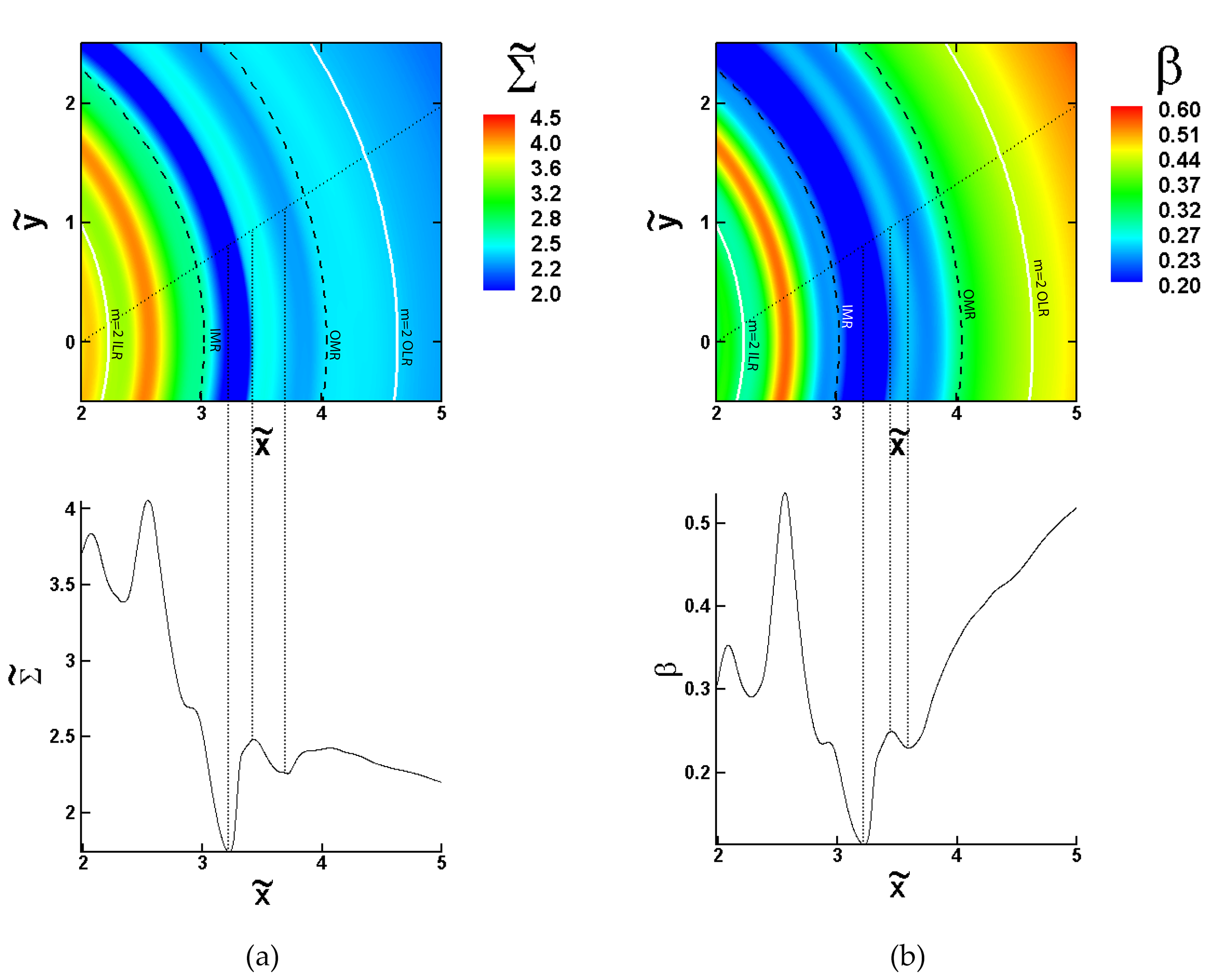}
  \caption{The (a) $\widetilde{\Sigma}$ and (b) $\beta$ distribution at $\widetilde{t} = 600$ for $\widetilde{\mathfrak{B}}_{\varphi} = \slfrac{6}{\widetilde{r}}$ 
   for a Jupiter-mass planet in region B indicated in \fig{fig:jup-mhd-total}. The IMR, OMR, $m=2$ OLR, and $m=2$ ILR are 
   labeled, as well as the location of the planet, in the top panel. $\widetilde{\Sigma}$ is sampled along a scanline indicated by a straight dotted line in the top panel and 
   plotted as a one-dimensional profile in the bottom panel of both (a) and (b).Vertical lines connecting the top and bottom panels 
   highlight variations in $\widetilde{\Sigma}$ and $\beta$ between the magnetic resonances and near the planet's orbital radius.}
  \label{fig:jup-mhd-b}
\end{figure*}

\fig{fig:jup-mhd-total} shows the density distribution at $\widetilde{t} = 600$ for the MHD model with  
$\widetilde{\mathfrak{B}}_{\varphi} = \slfrac{6}{\widetilde{r}}$ for the case of a Jupiter-mass planet.
The one-armed spiral wave associated with the $m=1$ OLR is also seen here, but it appears 
relatively more ``smeared out'' than in the hydrodynamic case. 
The two indicated regions in \fig{fig:jup-mhd-total}, A and B, are examined in more detail 
in Figs. \ref{fig:jup-mhd-a} and \ref{fig:jup-mhd-b}. Region A was chosen to show the disc properties very near 
the planet, while region B was chosen to show the disc properties near the planet's orbital radius but not near the planet
itself. In all of the following plots in this section, the locations of the $m=1$ OLR, the $m=2$ ILR and OLR,
the IMR and OMR, and the planet are labeled.

Figures \ref{fig:jup-mhd-a} and \ref{fig:jup-mhd-b} show the $\widetilde{\Sigma}$ and $\beta$ distributions in 
regions A and B, respectively. In each of these plots, $\widetilde{\Sigma}$ or $\beta$ is sampled
along a scanline; the scanline is indicated by a straight dotted line in the contour plot. The one-dimensional profile of the
quantity sampled along the scanline ($\widetilde{\Sigma}$ or $\beta$) is shown below the contour plot. The x-axis of both plots
are the same and are lined up such that the two plots can be directly compared and variations in $\widetilde{\Sigma}$ or $\beta$
can be more easily shown. 

In \fig{fig:jup-mhd-a}, the surface density is relatively higher at the planet's location than in the region
between the planet and the magnetic resonances. This is more clearly seen in the one-dimensional profile of
$\widetilde{\Sigma}$ sampled along the indicated scanline. Even with the non-constant underlying surface density distribution in the disc, the ``dips''
in surface density are visible. 

These relatively low-density areas are more easily seen in region B (\fig{fig:jup-mhd-b}), 
without the planet nearby. In region B, the relatively high density area corresponds to a ring of material located 
approximately at the planet's orbital semimajor axis that is seen on a larger scale in \fig{fig:jup-mhd-total}. So, 
there appears to be a ``ring'' of material near the planet's orbital radius, with 
``dips'' in $\widetilde{\Sigma}$ on either side of the planet's orbital radius between the magnetic resonances.

In Figs. \ref{fig:jup-mhd-a} and \ref{fig:jup-mhd-b}, $\beta$ is also relatively high near the planet's orbital radius, 
and it is relatively lower elsewhere between the magnetic resonances. 
The ``dips'' in $\beta$ correspond to a  relatively higher magnetic pressure than matter pressure in these regions. 
This correspondence of lower $\beta$ in regions of lower $\widetilde{\Sigma}$ indicates that there is an underdensity in regions with
relatively high magnetic pressure, as expected. 

\begin{figure*}
  \centering
  \includegraphics[width=\textwidth]{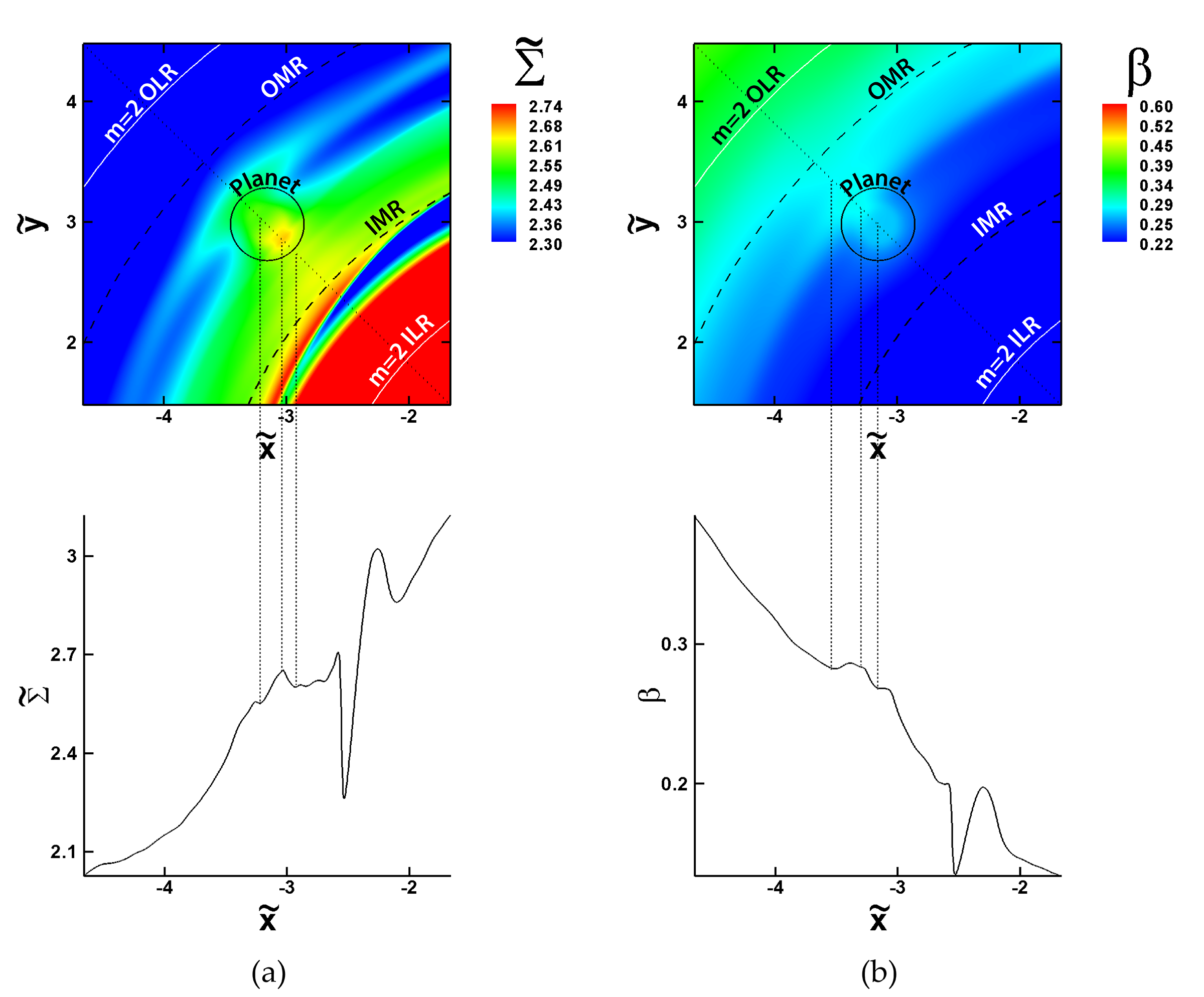}
  \caption{The (a) $\widetilde{\Sigma}$ and (b) $\beta$ distribution at $\widetilde{t} = 420$ for $\widetilde{\mathfrak{B}}_{\varphi} = \slfrac{8}{\widetilde{r}}$ 
   near the planet for a Saturn-mass planet in region A indicated in \fig{fig:jup-mhd-total}. The IMR, OMR, $m=2$ OLR, and $m=2$ ILR are 
   labeled, as well as the location of the planet, in the top panel. $\widetilde{\Sigma}$ is sampled along a scanline indicated by a straight dotted line in the top panel and 
   plotted as a one-dimensional profile in the bottom panel of both (a) and (b). Vertical lines connecting the top and bottom panels 
   highlight variations in $\widetilde{\Sigma}$ and $\beta$ between the magnetic resonances and near the planet's orbital radius.}
  \label{fig:sat-mhd-a}
\end{figure*}

\begin{figure*}
  \centering
  \includegraphics[width=\textwidth]{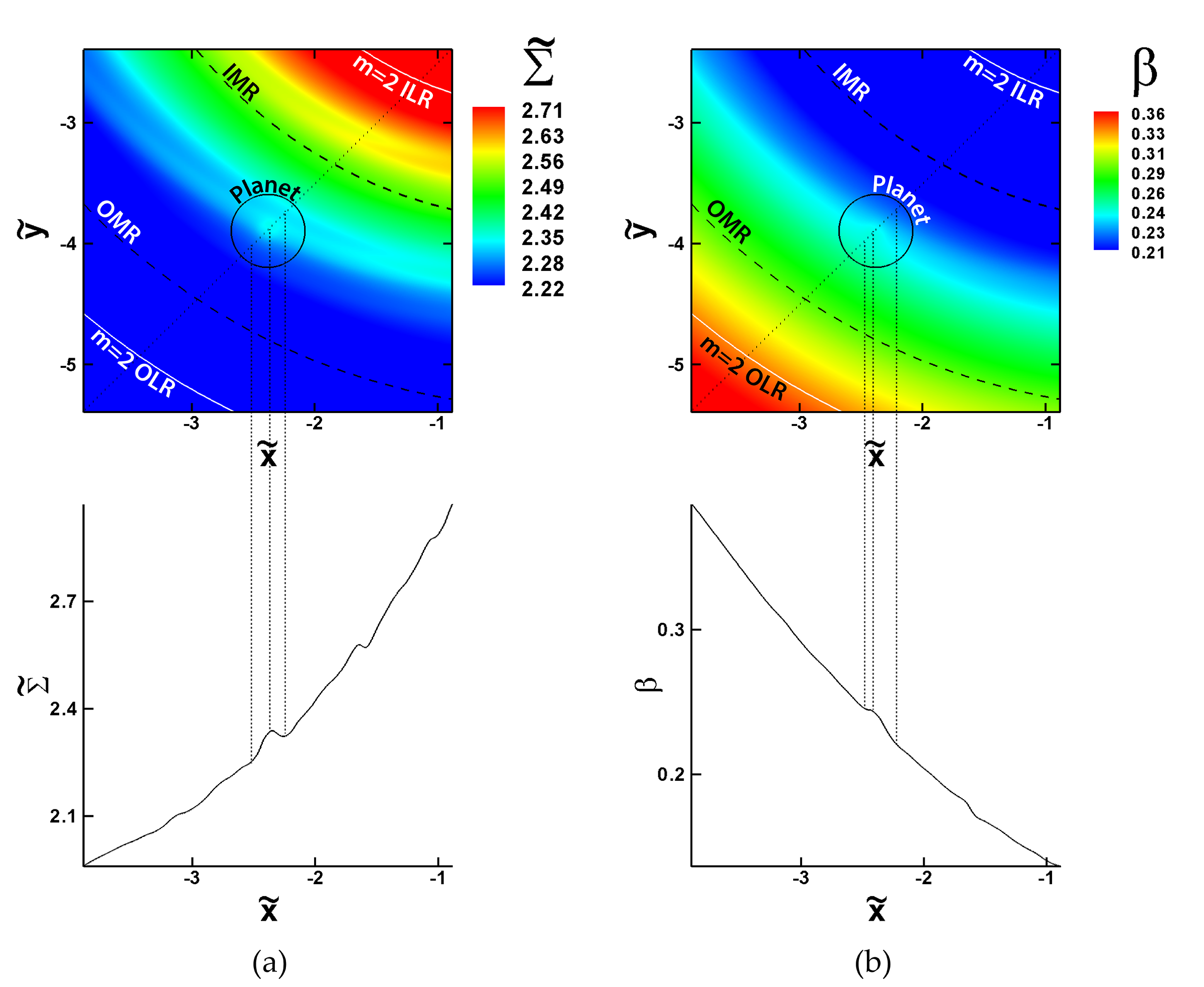}
  \caption{The (a) $\widetilde{\Sigma}$ and (b) $\beta$ distribution at $\widetilde{t} = 300$ for $\widetilde{\mathfrak{B}}_{\varphi} = \slfrac{8}{\widetilde{r}}$ 
   near the planet for a $5 M_{\oplus}$ planet in region A indicated in \fig{fig:jup-mhd-total}. The IMR, OMR, $m=2$ OLR, and $m=2$ ILR are 
   labeled, as well as the location of the planet, in the top panel. $\widetilde{\Sigma}$ is sampled along a scanline indicated by a straight dotted line in the top panel and 
   plotted as a one-dimensional profile in the bottom panel of both (a) and (b). Vertical lines connecting the top and bottom panels 
   highlight variations in $\widetilde{\Sigma}$ and $\beta$ between the magnetic resonances and near the planet's orbital radius.}
  \label{fig:5earth-mhd-a}
\end{figure*}

Figure \ref{fig:sat-mhd-a} shows similar $\widetilde{\Sigma}$ and $\beta$ contours and
one-dimensional profiles sampled along the scanlines for a Saturn-mass planet, while Figure \ref{fig:5earth-mhd-a} shows 
the same information for a $5 M_{\oplus}$ planet. Only the region nearby the planet is shown for the Saturn-mass and $5 M_{\oplus}$ cases, as
the perturbation to the disc from these smaller masses is small enough that the perturbations due to the MHD waves are much
harder to resolve far away from the planet. The overdensity near the planet, with underdensities on the opposite sides of the planet, within the magnetic resonances is
visible for both the Saturn-mass and $5 M_{\oplus}$ planets, as it is for the Jupiter-mass planet. The magnitudes of the
variations in $\widetilde{\Sigma}$ and $\beta$ are less visible for the Saturn-mass planet than the Jupiter-mass planet, 
and the variations are even smaller for the $5 M_{\oplus}$ planet. 

Overall, the magnetic resonances appear to alter the disc such that there is an underdensity (and relatively low value of $\beta$) within
the magnetic resonances, while there is an overdensity (and relatively high value of $\beta$) near the planet's orbital radius.

%%% Jupiter mass slices %%%

%%% Saturn mass slices %%%

%%% 5 Earth mass slices %%%

%%%%%%

\section{Torque on the Jupiter-mass planet}\label{section:torque_changes}

\begin{figure*}
  \centering
  \includegraphics[width=\textwidth]{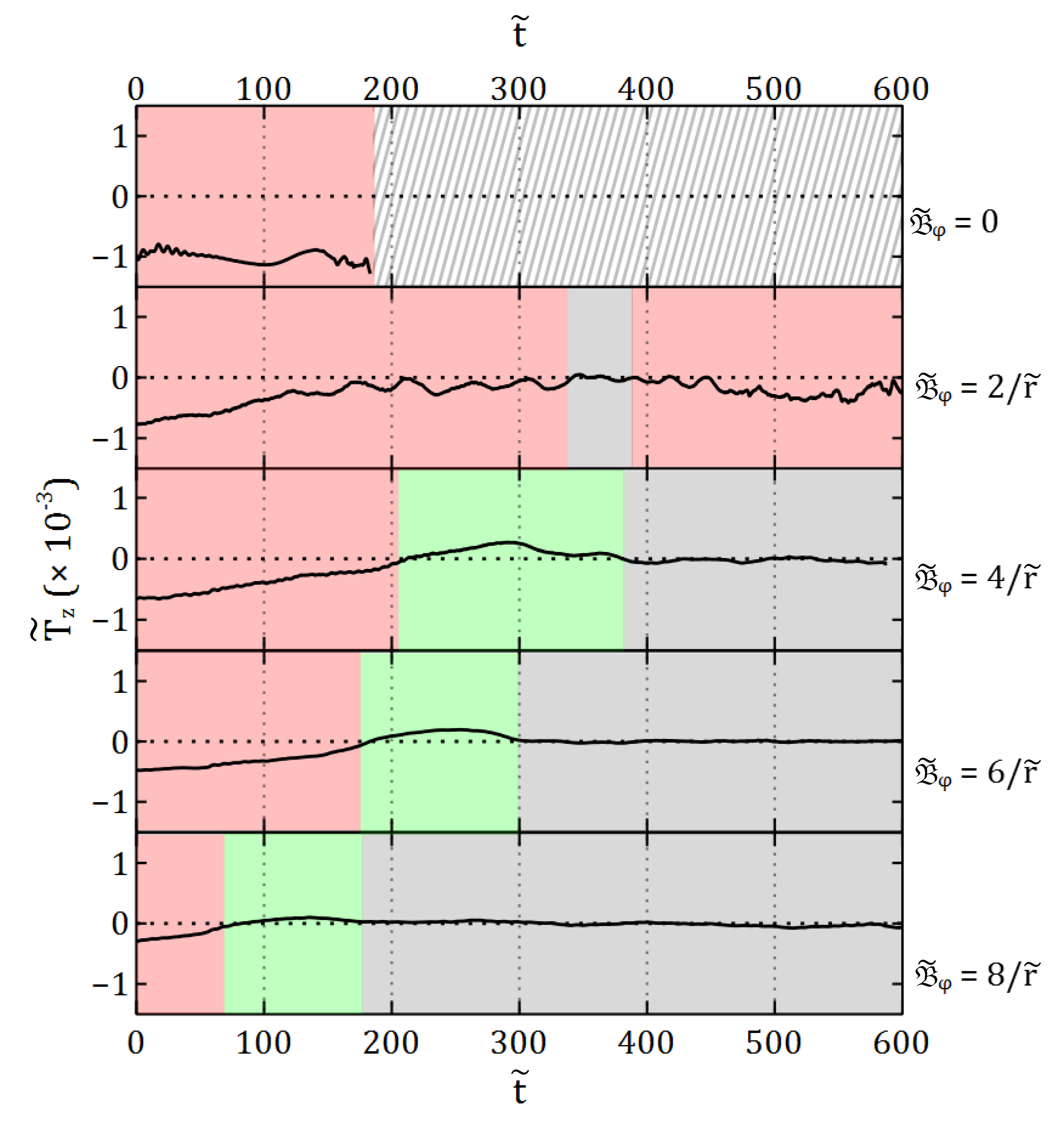}
  \caption{The change in the torque in the $z$ direction on the Jupiter-mass planet from the
   disc for several magnetic field amplitudes (labeled above). The line at zero torque separates negative and 
   positive torque. positive torque results in outward migration (shaded green) and negative torque results in inward migration (shaded red); zero
   torque represents a saturated torque, of which no migration is a consequence (shaded gray). The hatched region for the hydrodynamic
    case indicates that there is no data at those times, as the planet had already migrated beyond the inner boundary of the simulation region.}
  \label{fig:torque_vs_time}
\end{figure*}

\fig{fig:torque_vs_time} shows the smoothed\footnote{We smoothed the torques using iterative relaxation \citep{atp1984,ik1966}.} 
torque on a Jupiter-mass planet as a function of time, for several different magnetic field amplitudes. 
If a region is shaded red, the torque on the planet is negative and the migration is directed inward; if a region is 
shaded green, the torque is positive and the migration is directed outward; if a region
is shaded gray, the torque is roughly zero and the migration is ``stalled'' (likely because the torque has saturated); and the hatched region in
the hydrodynamic case indicates that the planet is no longer in the simulation region beyond that
time. This figure shows that a larger magnetic field amplitude corresponds to a larger total positive torque on a Jupiter-mass planet that 
happens earlier, faster, and over a shorter period of time. This also implies that the torque is saturated earlier for a larger magnetic field 
amplitude for a Jupiter-mass planet. 

So, for relatively low up to relatively high magnetic field amplitude, the torque on the Jupiter-mass planet is increasingly positive,
and this torque saturates at an increasingly large orbital radius at an earlier time.
This positive torque followed by saturation effectively stifles the inward differential Lindblad torque earlier for larger magnetic field amplitudes.

\section{Change in the planet's orbital semimajor axis}\label{section:semimajor_axis_change}

\begin{figure*}
  \centering
  \includegraphics[width=\textwidth]{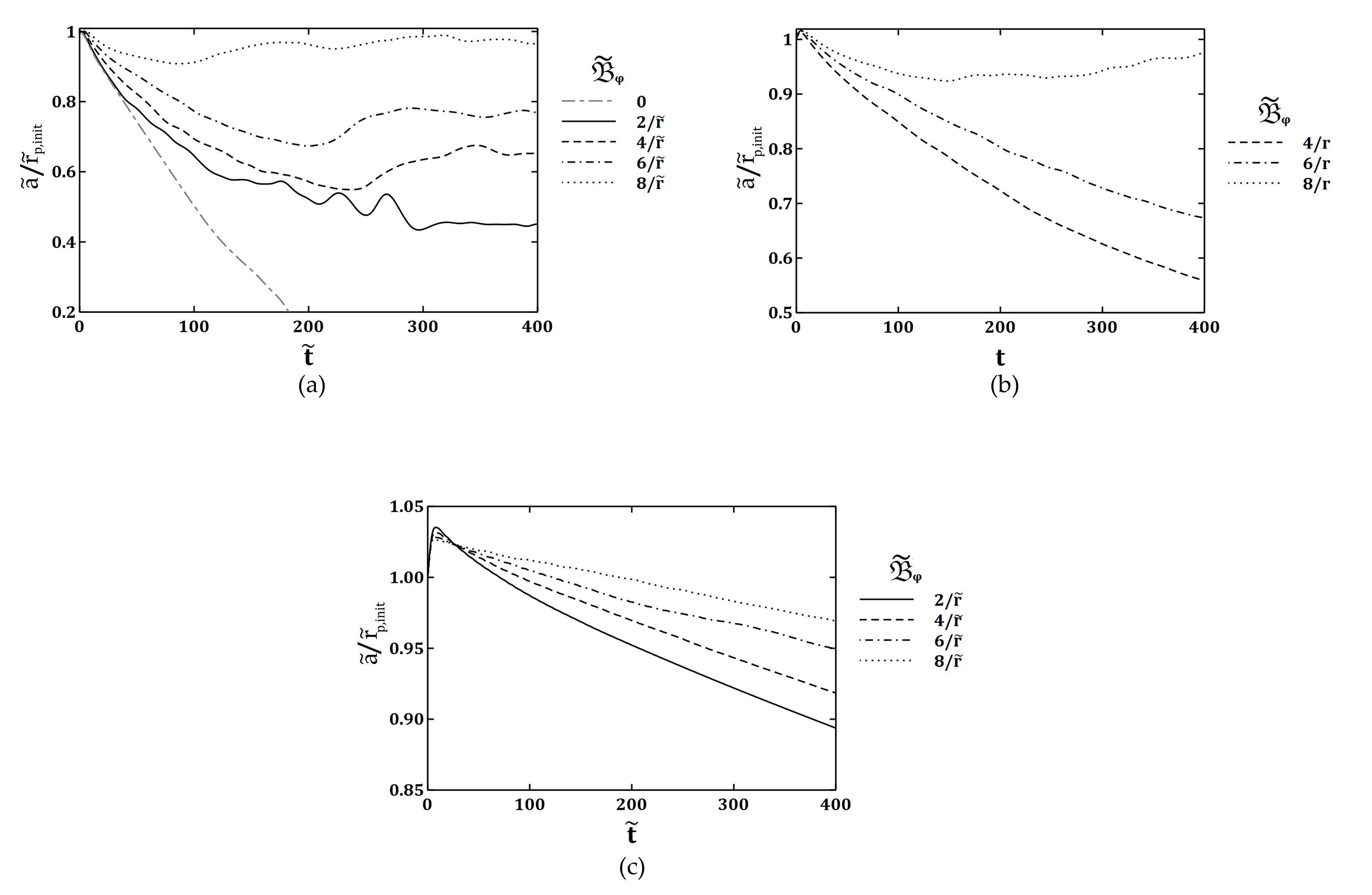}
  \caption{The normalized change in the planet's orbital semimajor axis over time. Panel (a) shows this change for a 
   Jupiter-mass planet, panel (b) shows this change for a Saturn-mass planet, and panel (c) shows this change for a
   $5 M_{\oplus}$ planet.}
  \label{fig:sax_vs_time}
\end{figure*}

Figure \ref{fig:sax_vs_time} shows the normalized change in the planet's orbital
semimajor axis over time for a Jupiter-mass, Saturn-mass, and $5 M_{\oplus}$ planet, respectively. 
Several magnetic field amplitudes are shown.

For the Jupiter-mass planet, the hydrodynamic case ($\widetilde{\mathfrak{B}}_{\varphi}~=~0$) results in rapid inward migration, as expected; 
the planet's inward movement is slowed in all of the MHD cases. For relatively low magnetic field amplitudes, the Jupiter-mass planet's migration remains inward.
For larger magnetic field amplitudes, however,  $\beta$ is larger near and interior to the planet, 
and its migration slows as a result. The migration is even slightly reversed for the largest magnetic field
amplitudes tested in this work. 

The initial inward migration becomes slower and takes place over a shorter period of time for larger magnetic field amplitudes. The ensuing
outward migration also slows, but takes place over a slightly longer period of time, for relatively low up to relatively high for magnetic field amplitude. 
So, for larger magnetic field amplitudes: the Jupiter-mass planet's inward migration is
slower than in cases of lower magnetic field amplitude, and the innermost radius reached by the planet increases.

In the case of a Saturn-mass planet, the migration is also directed inward in general, but it is slowed when the magnetic field amplitude is
increased. The Saturn-mass planet's migration also reverses for the largest field amplitude similarly to the Jupiter-mass planet.
For the $5 M_{\oplus}$ planet, the migration does not reverse, but it is slightly slowed when the magnetic field amplitude is
increased. The overall change in semimajor axis of the $5 M_{\oplus}$ planet is small relative to the change seen for the
Jupiter-mass and Saturn-mass planets, but there is still a noticeable change in its migration behavior.

\section{Dimensional Units}\label{section:dimensional_units}

Using Table \ref{table:units}, we can convert our results to dimensional units. 
Not reflected in Table \ref{table:units}, however, is the fact that we used a much higher
surface density in the disc than is considered realistic. 
The migration time of a planet depends inversely on the surface density of the disc 
\citep{ttw2002}~\footnote{This is strictly true when the surface density is a power law. Our surface density is not
an exact power law, but is very similar. (See Figure \ref{fig:initial_1d_sigma}.)}:
\begin{equation}
	t_{\rm mig} \propto \frac{M_{\star}^{2}}{M_{\rm p} \Sigma(r_{\rm p}) r_{\rm p}^{2} \Omega_{\rm p}} \left(\frac{c}{r_{\rm p} \Omega_{\rm p}}\right)^{2}.
\end{equation}
By using a very large surface density, our simulations require much less computational time to show appreciable planet migration. 
A typical protoplanetary disc surface density is $\Sigma \sim 10$ g cm $^{-2}$ \citep[e.g.,][]{wc2011}, while our surface density is $\Sigma \sim 10^{4}$ g cm$^{-2}$.
So, the migration times in our simulations must be multiplied by a factor of $1000$ to attain a migration time relevant to realistic physical systems.

Thus, in order to compare our work to realistic protoplanetary systems, we must alter our reference values for $\Sigma_{0}$, $B_{0}$, and $P_{0}$ as well:
$\Sigma_{0}' = \slfrac{\Sigma_{0}}{1000} = 35.5$ g cm$^{-2}$, $B_{0}' = \slfrac{B_{0}}{\sqrt{1000}}	= 9.2$ G, and $P_{0}' = 1000P_{0} = 350$ years.
We use the value of $\widetilde{a}$ from Figure \ref{fig:sax_vs_time}
to calculate the dimensional value of the planet's semimajor axis. For example, at $t = 0$ for a Jupiter-mass planet with $\widetilde{\mathfrak{B}}_{\varphi} = \slfrac{6}{\widetilde{r}}$,
$r_{\rm p} = 2.5$ AU, $\Sigma(r_{\rm p}) = 79$ g cm$^{-2}$, and $B_{\varphi}(r_{\rm p}) = 11$ G.

That the migration time scale for a planet is larger (i.e., the average migration rate is slower) 
when there is a larger magnetic field amplitude in the disc is evident in \fig{fig:sax_vs_time}.
For example, consider the Jupiter-mass planet whose orbital radius is initially $2.5$ AU. When this planet is embedded in a
non-magnetic disc, its orbit decreases to $0.5$ AU in approximately $8\times 10^{4}$ years. However, when this planet is
embedded in a disc with a magnetic field varying as $\widetilde{\mathfrak{B}}_{\varphi} = \slfrac{6}{\widetilde{r}}$, its orbit is only
reduced to $1.7$ AU in the same amount of time.

A more detailed description of the migration rates can be done for the case of a $5 M_{\oplus}$ planet, because the migration
direction does not change (see Table \ref{table:mig_rate}). From this information, it can be estimated that, for example,
a $5 M_{\oplus}$ planet will reach the surface of a $2R_{\odot}$ star after approximately $2.6$ Myr when
$\widetilde{\mathfrak{B}}(\widetilde{r})=\slfrac{2}{\widetilde{r}}$, while it would reach the surface of the star after approximately $9$ Myr when
$\widetilde{\mathfrak{B}}(\widetilde{r})=\slfrac{8}{\widetilde{r}}$.

\begin{table}
\begin{tabular} {c c c}
\hline
\multirow{2}{*}{$\widetilde{\mathfrak{B}}(\widetilde{r})$}		& Migration rate      			&  Migration time to \\
                                                                				& (AU yr$^{-1}$)      		& surface of star (Myr) \\
\hline
$\slfrac{2}{\widetilde{r}}$						&	$1.9\times 10^{-6}$	&	2.6\\
$\slfrac{4}{\widetilde{r}}$						&	$1.5\times 10^{-6}$	&	3.4\\
$\slfrac{6}{\widetilde{r}}$						&	$9.0\times 10^{-7}$	&	5.5\\
$\slfrac{8}{\widetilde{r}}$						&	$5.5\times 10^{-7}$	&	9.0\\
\hline
\end{tabular}
\caption{Approximate average migration rates for a $5 M_{\oplus}$ planet (corresponding to the semimajor axis changes shown in 
panel (c) of Figure \ref{fig:sax_vs_time}), and the approximate times until the planet reaches the surface of a $2R_{\odot}$ star based
on each of these rates.}
\label{table:mig_rate}
\end{table}

\section{Discussion and Conclusions}\label{section:conclusions}

The main results of our paper are:
\begin{enumerate}
\item{In the case of not only a Jupiter-mass planet, but also Saturn-mass and $5 M_{\oplus}$ planets, 
the magnetic resonances appear to alter the disc such that there is an underdensity (and relatively low value of $\beta$) within
the magnetic resonances, with an overdensity (and relatively high value of $\beta$)
at the planet's orbital radius. 
The magnitudes of the variations in $\Sigma$ and $\beta$ are smaller in the case of a Saturn-mass planet than for the Jupiter-mass planet,
and are still smaller in the case of a $5 M_{\oplus}$ planet, but the variations are still visible relative to the background
$\Sigma$ and $\beta$ distributions.
This agrees with results presented in \citet{terquem2003} and \citet{ftn2005} in which a 
$5 M_{\oplus}$ planet on a fixed circular orbit experiences a torque from the nearby magnetic resonances, and
the behavior of $\Sigma$ and $\beta$ near the planet is similar.}
\item{For relatively low up to relatively high magnetic field amplitudes, there is an increasingly strong positive torque on the Jupiter-mass planet that
happens earlier, with a corresponding torque saturation at earlier times, effectively reducing the
effectiveness of the inwardly-directed differential Lindblad torque.
The behavior of the net torque as a function of time, shown in \fig{fig:torque_vs_time} corresponds to the changes in the Jupiter-mass planet's semimajor axis shown
in \fig{fig:sax_vs_time}. }
\item{The planet's inward migration is slowed (and can even be reversed), and its 
orbit stabilizes at an increasingly large radius for a larger magnetic field amplitude. The migration both slowed and
reversed for the Jupiter-mass and Saturn-mass planets, while it slowed but did not reverse for the $5 M_{\oplus}$ planet.
In our simulations, we kept the mass of the disc constant at $1 M_{\rm Jup}$, and thus the amplitudes of the excited
waves are much smaller for the lower mass planets that we tested. Thus, the $5 M_{\oplus}$ planet migrated much more slowly than
the Saturn-mass and Jupiter-mass planets and never reached the region of the disc in which the magnetic field is
strong enough to stop its migration.} 
\end{enumerate}

\section*{Acknowledgements}
We gratefully acknowledge support from the NASA Research Opportunities in Space and Earth Sciences 
(ROSES) Origins of Solar Systems grant NNX12AI85G. We also acknowledge support from grants
FAP-14.B37.21.0915, SS-1434.2012.2, and RFBR 12-01-00606-a.

We also thank J. C. Mergo for valuable insight and comments.

%-------------------------------------------------------------------------------
% Bibliography
\bibliographystyle{mn2e}
\bibliography{draft74}
\end{document}